\documentclass[a4paper]{article}

\usepackage[english]{babel}
\usepackage{amsmath}
\usepackage{graphicx}
\usepackage{fullpage}
\usepackage{enumerate}

\title{HAS   QUANTUM   KEY   DISTRIBUTION   BEEN   PROVED   SECURE?\footnote{This v.2 is a re-writing of v.1 for a broader group of readers. It also contains a few new points.}}

\author{Horace P. Yuen\\Department of Electrical Engineering and Computer Science\\Department of Physics and Astronomy\\Northwestern University, Evanston Il. 60208\\email: yuen@eecs.northwestern.edu}

\date{}

\begin{document}
\linespread{1}
\maketitle

The term ``quantum key distribution" (QKD) impresses laymen and crypto experts alike that it involves the use of esoteric quantum effects, say of a single optical photon or quantum entangled photons, which are said to guarantee perfect or almost perfect encryption security [1,2] not possible otherwise. The following gives a general account on why this has not yet been established and why it may be difficult to do so.

It is important to observe that a cryptosystem may actually be secure, such as the widely used AES (Advanced Encryption Standard), without having been proved secure. The lack of security proof implies that a supposedly secure cryptosystem may one day be found not secure by an ``enemy", apart from human errors and other practical issues, which happened many times in the past. For a given mathematical representation of a cryptosystem, security can only be guaranteed by a logical proof. It cannot be demonstrated by experiments, in contrast to most other issues in science and engineering, if only because there are an unlimited number of attack possibilities consistent with the mathematical model that can only be dealt with totally by mathematics and its connections with reality. There is the additional very serious issue whether the cryptosystem mathematical model captures the real world situation sufficiently. In fact, many QKD systems have been broken by detector blinding attacks [3] precisely because the single photon detectors used are not fully represented.
   
We will not discuss such modeling issues in this paper, and will focus on the foundational difficulties that QKD encounters in the proof of its information theoretic security, and in particular on what kind of security with what associated numerical security levels can be obtained. The possibility of such a security proof is what distinguishes QKD from other cryptosystems, despite its great inefficiency which is mainly due to single photon sources in QKD as compared to $>10^7$ photons in ordinary optical communications. We begin by a description of the basic security problem of privacy.

Protection of data privacy has become an increasingly important problem area that affects even our daily lives. A major problem concerns two users Alice and Bob who want to protect their private communications against an attacker Eve. Such privacy protection is commonly done by some form of encryption with a cipher that employs a prior shared key bit sequence $K$ known to both of the users. There are two main issues with this approach. How strong the cipher is against Eve's eavesdropping and other more active attacks, and how should $K$ be distributed between the users? If to Eve $K$ is the uniform random variable $U$, which has probability $2^{-n}$ for each of the $2^n$ $n$-bit sequences $K$ may take as value, and the bit length of the data sequence $X$ to be transmitted is also $n$, one can XOR each bit in the plaintext $X$ with a separate bit in $K$ to form the ciphertext $Y$, which is called ``one-time pad" encryption. It is ``perfectly secure" because while Bob can decrypt $X$ from $Y$ knowing $K$, Eve can not learn anything about $X$ from $Y$. The characteristics of such perfect $K=U$ is that the bit value 0 or 1 occurs with probability 1/2 in each bit and the bits are statistically independent. It has ``information theoretic security" because $X$ remains totally probabilistically uncertain to Eve from her attack, which may be contrasted with complexity based security in which $X$ cannot be found by Eve only because she is limited in computational power. The distinguishing character of one-time pad is that it is ``perfectly secure", its information theoretic security is at the highest possible level. This is because knowing $Y$ does not help Eve even probabilistically in recovering any bit in $X$. 

The problem of the one-time pad is that the required length of $K$ is far too long in almost all practical situations. In the usual approach, a cipher that employs a much shorter $K$ than $X$ is used but which has no provable security. The distribution of such short $K$ is often accomplished by a public key method which has only complexity based security that may succumb to future technology advances, in contrast to the information theoretic security of one-time pad, and moreover an additional unproved assumption is needed to ensure the complexity required for security. These problems are to be solved with QKD by letting Alice and Bob generate their own long key $K$ which is provably secure and then use it in the one-time pad form for encryption. 

\begin{figure}
\centering
\includegraphics[width=.75\textwidth]{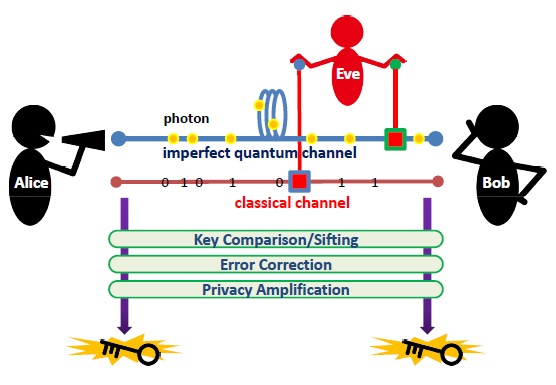}
\caption{Quantum Key Distribution, a typical protocol. Two users, Alice and Bob, communicate over a photon channel and a classical channel to establish a preliminary key, which is then corrected for error and compressed by privacy amplification (for increasing security level) to generate a final key. The attacker Eve can intercept both channels.}
\end{figure}

However, it is universally agreed that it is actually not possible to get a perfect key from a QKD protocol. In what sense then an imperfect key would guarantee the cryptosystem security? 

In the earlier days of QKD the criterion of ``accessible information" was used as the quantitative measure of security, which is Eve's maximum Shannon mutual information $I_E$ between the QKD generated key $K$ and the observation Eve makes from her attack on the cryptosystem. The $K$ was taken to be perfectly secure asymptotically when $I_E$ goes to zero as the number of bits $n$ in $K$ goes to infinity, and $n$ becomes the ``security parameter" of the cryptosystem, meaning that the security can be made arbitrarily close to perfect by making the security parameter arbitrarily large.  

The key question that has \textit{not} been fully resolved to date is the following. Given a security criterion $S_c$ such as $I_E$ that is equal to 0 for perfect security, what is the security guarantee when $S_c>0$ is at some quantitative value? On the other hand, enough has been found about two common criteria, the $I_E$ above and the trace distance criterion $d$ to be discussed, that some serious problems on security guarantee can be pinpointed.

Generally, in an attack Eve could obtain a probability distribution $P(K)$ on the possible values of $K$, and the quality of an imperfect $K$ needs to be assessed by comparing $P(K)$ to the uniform distribution $U$. Any security criterion $S_c$ consisting of a single number does not capture the full distribution function $P(K)$, the details of which relate to various possible security breaches. One most important security measure from $P(K)$ is $p_1$, Eve's maximum probability of identifying the whole $K$ from her attack. Clearly $p_1$ has to be sufficiently small. Once $p_1$ is brought into the security picture, it becomes evident [4] it is the rate that $I_E$ goes to 0, not the fact that it does, which determines quantitative security level at any key generation rate. It turns out there is always a tradeoff between security level and key rate, with lower security for higher key rate, and there is no security parameter in QKD with which the security level can be driven to perfect [4, 5].

In this connection, it is important to note that the various entropies and mutual information are not operational quantities and may become misleading. ``Information" here is a technical term with different meaning from its daily usage, as Shannon himself emphasized in the 1950's. In ordinary communications, their operational meaning is given by relating them to the empirical data rate and error rate through the Shannon coding theorems. In cryptography the operational meaning of various criteria $S_c$ were spelled out in part only recently [4,5].

It turns out that accessible information is not a good quantitative security criterion in QKD due to the phenomenon of quantum information locking, and it has been replaced by a quantum trace distance criterion $d$ [1, 2] by the majority of QKD groups. We do not need to be concerned with the detailed definition or technical situation for the purpose of this paper, and may just note that $0\leq d\leq 1$ and the smaller $d$ is the closer it is to the ideal $d=0$ case. There are several problems associated with $d$ and its use in QKD protocols. 

(1) The trace distance criterion $d$ is commonly interpreted as the maximum ``failure probability" of the QKD system, ``failure" meaning the system is not ``ideal" [1,2]. A key from a system not ideal is also imperfect. Thus $d$ is taken to be the (maximum) probability that the key is not perfect, not being $U$ to Eve. It turns out that under a $d$ guarantee, $p_1$ can be as large as $d+2^{-n}$ for an $n$-bit $K$ [4], meaning the whole $K$ may be revealed to Eve with a probability (slightly bigger than) $d$. A 1,000 bit  perfect $K$ would give probability $2^{-1,000}$ for $p_1$, while the best value in theory [6] has $d>2^{-50}$. This is hundreds of orders of magnitude away from a perfect key. Experimentally $d\sim 10^{-9}$ for a $10^5$ bits $K$ [7] and the gap from perfect is even more enormous.

This interpretation is actually incorrect [4,5,8], and the probability of $K$ being imperfect is 1, not $d$. However, bounds on Eve's average probability of successfully estimating various portions of $K$ can be obtained [5] in terms of $d$. The implications of this error are discussed in the following points (2)-(5).

(2) The $d$ guarantee obtained for QKD protocols is an average value over the very large number of possible privacy amplification codes used in generating $K$. (``Privacy amplification" involves compressing or ``hashing" a number of given bits into a smaller number for increasing the security level, which is a necessary step in QKD protocols.) For product guarantee one does not use average guarantee but uses probability guarantee, which can be obtained from average for the unknown $P(K)$ through the use of Markov inequality. The necessity of such conversion is easily seen from the (artificial) example of a cryptosystem which has 50/50 chance of being secure for 100 years or totally insecure, with an average of 50 years of security. When a single average guarantee over privacy amplification is converted into probability guarantee, the minimum ``failure probability" is increased from $d$ to $d^{1/2}$. In addition, such privacy amplification guarantee requires a $p_1$ guarantee on the hashing input bits. Only averaged $p_1$ over Eve's possible attack measurement results is prescribed in security proofs. To remove such average, Markov inequality needs to be used also for a different random quantity. The minimum failure probability that can be obtained then becomes $\sim d^{1/3}$ [5].       

In addition to ciphertext-only attack in which Eve knows only the ciphertext, Eve may launch a \textit{known-plaintext attack} [9] when the generated $K$ is used for one-time pad encryption, in which Eve  also knows some bits in the plaintext $X$. Since the whole ciphertext $Y$ is always taken to be open, Eve would know the corresponding bits in $K$. For one-time pad encryption this does not help her to get other unknown bits in $X$ she doesn't know, because the bits in $K$ are independent. However, for an imperfect $K$ there may be correlation among its bits that would let Eve learn about the other bits in $K$, and hence the other bits in $X$ through $Y$. It is easy to construct specific example in which knowing some specific $m$ bits of $X$ would reveal all the rest of $X$ with probability 1 from such attack, however long $X$ may be, when $d=2^{-m}$. This, incidentally, is a specific counter-example to the failure probability interpretation of $d$. There is an average guarantee over the known bits of $X$ in terms of $d$. After this additional average is also removed, the correct failure probability guarantee becomes $d^{1/4}$ [5].

(3) From point (2) above the numerical values of the probability guarantee get much worse than even those of point (1). The theoretical and experimental values become $d\sim 10^{-3.5}$ and $d\sim 10^{-2.3}$. They are clearly not meaningful security guarantee for even a single QKD round. Typically 10 QKD rounds are carried out in a sec, and thus $\sim10^6$ rounds per day. When the $d$ value as failure probability is converted to a daily rate, the possibility of Eve totally breaking about $10^3$ rounds per day is not ruled out by the security guarantee.

If a probability of $10^{-15}$ for a single trial is to be considered practically impossible and there are just $10^5$ QKD rounds total to be carried out, $d$ would need to be $10^{-80}$ for the effective $d^{1/4}$ probability to be $10^{-20}$ needed for security guarantee against known-plaintext attack, and $d=10^{-60}$ is needed for ciphertext-only attack. Compare this to the best theory value of $10^{-14}$. It may be mentioned that for a ciphertext-only attack the obtainable QKD security level is quite unfavorable compared to the information theoretic security that can be obtained from symmetric key expansion in conventional cryptography [8]. The latter has, however, no information theoretic security against know-plaintext attack.

(4) When Eve makes an error in estimating $K$ or one of its subsets, she may nevertheless get many bits correctly since one bit error already makes the whole sequence in error. This is the analog of the ``bit error rate" issue in communications, which is different from ``sequence error rate". For a perfect key the guarantee remains perfect, but under a $d$ guarantee it is not yet quantified. Similar deviation from an imperfect key may allow ``distinguishing attack" [9] that has not yet been discovered.  

(5) Quantification of security with error correction, which is a necessary step in QKD with small energy signals, is an almost insurmountable problem. For a given error correction procedure, it is not possible to rigorously estimate how many errors Eve could correct with what probability, from her knowledge of any publicly announced error correction procedure. Error correcting codes are notorious for their capability of correcting many more errors than those theoretically predicted. No justification has been given for the widely used $leak_{\text{EC}}$ formula for such information leak to Eve [5]. Even if the parity check digits of a linear error correcting code are covered by true one-time pad, there is still the leak of code structural information. And with an imperfect key used to cover such digits the resulting security level is uncertain, especially at the relatively large $d$ levels that can be obtained. The seriousness of the problem can be seen from the use of imperfect key in message authentication, in which the tag length security parameter is lost due to the imperfect key [5].

The foundational consideration pointed out in this paper applies to all QKD protocols, including the recent measurement-device-independent approach. It shows that adequate security for QKD protocols has not yet been obtained just in theory, and it does not seem possible to obtain it without some new ingredient and further research. There are also many other approach specific security problems that have not been resolved [10]. The trace distance criterion is further formally treated recently [11], with no mention of any of the long standing difficulties indicated in this paper which are not resolved as before. We end this note by quoting [9, p19] that it is important to ``...realize that security is incredibly subtle and that it is very easy to overlook critical weaknesses."

\begin{center}
\textbf{Acknowledgment}
\end{center}
I would like to thank D. Abbott, O. Hirota, L. Kish, and especially G. Kanter for many useful discussions on cryptography and suggestions on how to make some parts of cryptography more readily understandable.

\end{document}